# High-temperature oxygen non-stoichiometry, conductivity and structure in strontium-rich nickelates $La_{2-x}Sr_xNiO_{4-\delta}$ (x = 1 and 1.4)

**L.V. Makhnach [a], V.V. Pankov [a], P. Strobel [b]**

[a] *Institute of General and Inorganic Chemistry, National Academy of Sciences of Belarus, 9 Surganova street, 220076 Minsk, Belarus*

[b] *Institut Néel, Dept. MCMF, CNRS, BP166, 38042 Grenoble Cedex 9, France*

## Abstract

Oxygen nonstoichiometry, electrical conductivity and thermal expansion of $La_{2-x}Sr_xNiO_{4-\delta}$ phases with high levels of strontium substitution ($1 \leq x \leq 1.4$) have been investigated in air and oxygen atmosphere in the temperature range 20-1050 °C. These phases retain the $K_2NiF_4$-type structure of $La_2NiO_4$ (tetragonal, space group I4/mmm). The oxygen vacancy fraction was determined independently from thermogravimetric and neutron diffraction experiments, and is found to increase considerably on heating. The electrical resistivity, thermal expansion and cell parameters with temperature show peculiar variations with temperature, and differ notably from $La_2NiO_{4\pm\delta}$ in this respect. These variations are tentatively correlated with the evolution of nickel oxidation state, which crosses from a $Ni^{3+}/Ni^{4+}$ to a $Ni^{2+}/Ni^{3+}$ equilibrium on heating.

*Corresponding author*: *P. STROBEL*
Institut Néel, CNRS, case F, BP 166, F-38042, Grenoble Cedex 9, France
phone: 33+ 4 76 88 79 40
fax: 33+ 4 76 88 10 38
e-mail: pierre.strobel@grenoble.cnrs.fr

## 1. Introduction

La-Sr-Ni-O compounds with the $K_2NiF_4$-type structure have been widely investigated as possible materials for manufacturing of electrodes, resistors, oxygen-permeable membranes etc. They are also of fundamental interest because of their structural similarity with copper-based high-temperature superconductors. Most of the published articles devoted to the $La_{2-x}Sr_xNiO_{4\pm\delta}$ system deal with compositions with small strontium content, i.e. $x \leq 0.5$. The synthesis of strontium-rich compositions as single-phase is more difficult and such compounds have been much less studied. Works by Cava et al. [1] and Takeda et al. [2] reported single-phase solid solutions with x(Sr) as high as 1.6. The oxygen stoichiometry $La_2NiO_{4\pm\delta}$, i.e. the variation of δ with temperature and oxygen partial pressure has been studied in detail [3,4]. For $La_{2-x}Sr_xNiO_{4\pm\delta}$ with high x values, however, data remain very scarce. Cava et al. [1] assumed a stoichiometric oxygen content for $x > 0.75$. Studies by Sreedhar et al. [5] and Takeda et al. [2] determined oxygen stoichiometries of 3.95 and 3.96 for x = 1.3 and 1.6, respectively. The oxygen stoichiometry is obviously variable as a function of temperature and influences the electrical properties at high temperature, as shown by Vashook et al. by high-temperature oxygen coulometry [6]. This paper describes an investigation of the oxygen content, structural and electrical properties of strontium-rich compositions $La_{2-x}Sr_xNiO_{4\pm\delta}$ (x = 1 and 1.4) as a function of temperature in air and oxygen atmospheres.

## 2. Experimental

### *2.1. Synthesis*

Members of the $La_{2-x}Sr_xNiO_{4\pm\delta}$ family were prepared by both a standard ceramic route and a citrate process [7]. Starting chemicals were reagent-grade $La(NO_3)_3{\cdot}6H_2O$, $Sr(NO)_3$ and $Ni(NO_3)_3{\cdot}6H_2O$. For the ceramic route, the reagents in appropriate proportions were dissolved in distilled water and the solutions were dried on a water bath or slowly heated in a furnace up to 250 °C. The dried nitrate mixture was decomposed at 750-800 °C for 5-6 hours. The powder mixtures were then fired at 1250 °C in oxygen for 20 hours and cooled at 150°C/h.

For the citrate route, an aqueous solution of citric acid was added to the nitrate solution, using an amount of citric acid calculated for full nitrate-to-citrate conversion. The citrate-nitrate solutions were dehydrated in an IR-IM2 rotary evaporator at 70-100°C. Dried powders were heated slowly (25-40 °C/h) to 200 °C in a muffle furnace. The resulting powders were hydrostatically pressed at 400-600 MPa into pellets (diameter 10-15 mm, thickness 2-5 mm) and fired at 800-900 °C for 20-30 hours depending on composition. In both processes, four intermediate grindings were carried out in order to obtain single phase compounds. The conditions of the final sintering step are given in Table 1.

### *2.2. Structural characterization*

The intermediate and final products were characterized by X-ray diffraction (XRD) using DRON-3 (CrK$\alpha$ radiation) and Philips X`Pert MPD (CuK$\alpha$ radiation) diffractometers. In-situ X-Ray investigations were carried out in the MPD diffractometer with a Buhler HDK 2,4 high-temperature attachment in the temperature range 20-1000 °C in air. Crystal lattice parameters were refined by a least square method.

In order to obtain accurate data for oxygen occupation refinements, a $La_{0.6}Sr_{1.4}NiO_{4\pm\delta}$ sample was studied by neutron diffraction at Institut Laue-Langevin, Grenoble, France. The sample (ca. 10 g) was enclosed in an amorphous silica tube (diameter 12 mm) open to air and placed in a vertical furnace in the powder diffractometer D1B. Several diffraction patterns were recorded as a function of temperature with wavelength $\lambda$ = 1.911 Å. Structural parameters, including cell parameter, atomic coordinates, atomic displacement parameters and oxygen occupations were determined by Rietveld refinement using the Fullprof software [8].

### *2.3. Physico-chemical characterization*

Differential thermal analysis was performed using a Derivatograph-Q1500D connected with a thermogravimetric analyzer. Mass variations were recorded on heating and cooling using a 2.6 °C/min rate. The temperature of the samples was controlled to within 0.5 °C. The experiments were carried out in flowing air, oxygen, or hydrogen with a gas flow rate of 5 l/h.

The oxygen content of single-phase powders was determined independently by coulometric reduction of the samples in hydrogen and by iodometric titration. In the former procedure, a 50-100 mg sample was treated at 1050°C in hydrogen in an Oxylyt system (SensoTech, Magdeburg, Germany) [9]. This device allows to determine the amount of oxygen released in the reduction of nickel to metallic Ni, while $Sr^{2+}$ and $La^{3+}$ are not reduced. For iodometric titration, 50-80 mg of each sample was dissolved in a solution contained 10 ml of 1N potassium iodide and 10 ml of diluted (1:2) hydrochloric acid. The iodine formed was titrated against a standard 0.02 N solution of sodium thiosulfate.

Sample expansion on heating was determined using a quartz dilatometer DKV-4 (Russia) within the temperature range 20-950°C in air. Heating and cooling rates were 2.6°C/min.

Ceramic samples for electrical measurements were hydrostatically pressed at 400-600 MPa and sintered at 1000-1300 °C during 20 hours in air (x=0) or in oxygen flow (x>0). The electrical conductivity was measured using a standard DC four-point method on parallelipipedic samples ($10x3x3mm^3$).

## 3. Results

### *3.1. Synthesis and oxygen content*

Single-phase $K_2NiF_4$-type compounds were obtained only for x ≤ 1.4. Samples with higher strontium content were found to be multi-phase, independently of the synthesis method (ceramic or citrate route). XRD analysis showed that they consisted of a main $K_2NiF_4$-type nickelate phase, nickel oxide (NiO) and strontium hydroxide. The amount of impurities increased with increasing strontium content. The impurities could be eliminated by increasing neither the sintering temperature (to 1350 °C) nor the sintering time (up to 50 hours). Consequently, only the single-phase compositions (x = 1.0 - 1.4) were selected for the further investigations.

The results of chemical analysis on nickelates prepared in air are given in Table 1. The average oxidation state of nickel cations increases with increasing strontium content from +2.28 in $La_2NiO_{4.14}$ to +3.30 in $La_{0.6}Sr_{1.4}NiO_{3.95}$. The x = 1 composition is stoichiometric within experimental error, i.e. $LaSrNi^{3+}O_4$. Compositions with x > 1 are oxygen-deficient; the oxygen contents measured (see Table 1) are consistent with values reported previously for x = 1.3-1.6 [2, 5].

Thermogravimetric analysis (TGA) results for x = 1 and x = 1.4 in air and in oxygen are displayed in Fig.1. These measurements show that oxygen is reversibly released above ~520 °C, and this effect is more pronounced in air (curves 1,3) than in oxygen (curves 2, 4). The mass variations observed between room temperature and 1000°C in air correspond the a range of oxygen stoichiometry variation of 3.99 to 3.92 and 3.95 to 3.74 for x = 1 and x 1.4, respectively. Interestingly, the oxygen release from these quaternary phases begins at a temperature about 200°C higher than for $La_2NiO_{4\pm\delta}$ [3, 6].

### *3.2. Electrical conductivity measurements*

Figure 2 shows the variation of electrical resistivity of nickelates with three different strontium contents with temperature ρ(T) up to 1100°C. This graph includes data measured on heating and cooling at rate 2.6 °C/min in air (lines 1, 3) and in oxygen (lines 2, 4). At temperatures above ca. 400°C, the resistivities are always higher in air than in oxygen. This effect is much more pronounced for strontium-substituted samples than for $La_2NiO_{4\pm\delta}$ (upper panel). Regarding the variation of resistivity with temperature, all samples follow a metallic behaviour above 300°C with a weak positive slope in ρ(T) below 550 °C. At higher temperatures, the behaviour of strontium-substituted phases is again markedly different from that of $La_2NiO_{4\pm\delta}$. Whereas the resisitvity increases slightly and monotonously in the latter (Fig.2, upper curve), anomalies are clearly observed for the former : the resistance increases sharply to reach a maximum value near 800°C, then decreases again. The expected increase with temperature is restored at higher temperatures (850-1000°C). This feature is more promined for higher strontium content (curves 3 and 4 in Figure 2). In addition, this resisitivity "peak" is not observed on cooling. Finally, it should be pointed out that the coincidence of resistivity values before the experiment and after the heating/cooling cycle is much better for the measurement in oxygen (curves 2 and 4 in Fig.2) than in air (curves 1 and 3). In fact, the resistivity of $La_{0.6}Sr_{1.4}NiO_{4-\delta}$ was found to go back to initial value after 1 day storage in air at room temperature, showing that the kinetics of oxygen exchange reaction in the $K_2NiF_4$-type oxide structure is much faster at higher oxygen pressure.

### *3.3. Thermal expansion and cell parameters*

Thermal expansion data are shown in Fig. 3 for x = 0 ($La_2NiO_{4\pm\delta}$) and x = 1.4. These measurements were carried out using the same heating and cooling rate (2.6 °C/min) as for resistivity measurements. The dilatometric data for unsubstituted $La_2NiO_{4\pm\delta}$ increase monotonously with a slight slope increase around 350°C and no anomaly. For $La_{0.6}Sr_{1.4}NiO_{4-\delta}$ composition, on the contrary, abrupt slope decreases are observed at 520 and 810°C. These features correlate remarkably well with the anomalies observed in the variation of resistivity with temperature (see Fig.2). This measurement also clearly shows a discrepancy between the dilatometric curves recorded on heating and on cooling.

The dilatometric data were confirmed by in-situ high-temperature XRD data. For $La_2NiO_{4\pm\delta}$ (Fig.4a), both *a* and *c* cell parameters increase linearly with temperature over the 350-1000°C temperature range with almost parallel slopes. The evolution of cell parameters with temperature is dramatically modified in strontium-substituted samples. Firstly, the difference in slope between the variations of cell parameters *a* and *c* is much larger. Secondly, from ca. 600°C up, the value of $\Delta a/a_O$ levels off whereas that of $\Delta c/c_O$ increases more sharply.

### *3.4. Neutron diffraction*

The thermal evolution of cell parameters, of oxygen vacancies, and the possibility of oxygen vacancy ordering were also probed by neutron diffraction on the $La_{0.6}Sr_{1.4}NiO_{4-\delta}$ composition. The data were refined in the $K_2NiF_4$ structure type, space group I4/mmm, with the following atomic positions: (La,Sr) at 4e (0 0 z), Ni at 2a (0 0 0), O1 at 4c (1/2 0 0), O2 at 4e (0 0 z). The structural model is shown in Fig. 5. The results are summarized in Table 2, and a graphical example including $Y_{obs}$-$Y_{calc}$ data is given in Figure 6. The broad bumps at 2θ around 25 and 52° are due to the amorphous silica container. At T ≥ 700°C, a few weak extra reflections were observed (see Fig. 6 around 2θ = 42 and 110°). These are ascribed to the high-temperature form of cristobalite (cubic, Fd3m space group, a = 7.135 Å) and probably arise from slight recrystallization of silica. Including this phase in the refinements showed that its fraction does not exceed 1.0 % at 870°C.

The evolution of cell parameters from X-ray data is in excellent agreement with high-temperature X-ray data. Both the enhanced increase in *c* and the levelling off of *a* above 700°C are confirmed. The $La_2NiO_{4\pm\delta}$ system is known to give rise to vacancies on both lanthanum and oxygen sites at high temperature [4]. The refinement of occupation on the La/Sr site (4e) in $La_{0.6}Sr_{1.4}NiO_{4-\delta}$ yielded a negligible variation with respect to full occupation. On the contrary, an important increase in oxygen vacancy concentration as a function of temperature was found from refinements of neutron diffraction data (see Table 2). In addition, these refinements clearly showed that this oxygen non-stoichiometry is entirely concentrated on the O1 site, i.e. the equatorial oxygen atoms, which are located together with nickel atoms in the z = 0 plane (see Fig. 5). Tentative refinements of oxygen occupations on the O2 site (apical oxygen along c) always yielded values in the range 1.00–1.02 with standard deviations close to 0.02. The total number of oxygen atoms is $2(n_{O1} + n_{O2})$, so that the overall oxygen stoichiometry per formula unit from neutron data refinements is 3.96, 3.84 and 3.75 at 400, 700 and 870°C, respectively. The 400°C value is consistent with chemical analyses (see Table 1).

An examination of Ni-O bond lengths (Table 3) shows that the nickel octahedral site is elongated along c (2 longer Ni-O2 distances), in agreement with previous studies [2]; this distortion is slightly enhanced at higher temperature. The increase in average Ni-O interatomic distance at higher temperature is consistent with the increase in oxygen vacancy concentration, since the latter induces a decrease in nickel oxidation state, hence an increase in Ni-O bond length. Note that to the resolution of the neutron diffractometer used, we found no evidence of oxygen vacancy ordering such as that reported by Medarde et al. [10]. It must be pointed out that this feature, leading to an orthorhombic Immm structure with a splitting of the O1 site between O1 (at ½00) and O1' (at 0½0), was only observed at low temperature for $0.15 \leq x(Sr) \leq 0.5$. No superstructure was found in room-temperature diffraction patterns in the subsequent very comprehensive neutron diffraction study of Millburn et al. [11] for $0.2 \leq x(Sr) \leq 1$. No high-temperature neutron diffraction data are available for higher strontium contents, to our knowledge.

Oxygen occupation values in Table 2 show that neither the oxygen stoichiometry nor the *c*/*a* value are restored in $La_{0.6}Sr_{1.4}NiO_{4-\delta}$ on cooling down to 400°C, in spite of the rather large time scale of the neutron diffraction experiment (each temperature plateau lasted at least six hours). This is consistent with the hysteresis observed in the measurements of resistivity and thermal expansion (see Figs. 2 and 3).

## 4. Discussion

Thermogravimetric and neutron diffraction studies showed clearly (i) that the fraction of oxygen vacancies in $La_{2-x}Sr_xNiO_{4-\delta}$ (x = 1 or 1.4) increases considerably with temperature and is enhanced at lower partial oxygen pressure, (ii) that these vacancies are located in the nickel planes. The release of oxygen can be expressed by reactions such as :

$$LaSrNi^{+2.98}O_{3.99}\square_{0.01} \Leftrightarrow LaSrNi^{+(2.98+2\delta)}O_{3.99-\delta}\square_{0.01+\delta} + (\delta/2)O_2 \quad (1)$$

$$La_{0.6}Sr_{1.4}Ni^{+3.30}O_{3.95}\square_{0.05} \Leftrightarrow La_{0.6}Sr_{1.4}Ni^{+(3.30+2\delta)}O_{3.95-\delta}\square_{0.05+\delta} + (\delta/2)O_2 \quad (2)$$

This situation is rather different from that in $La_2NiO_{4\pm\delta}$ case, where the initial stoichiometry includes interstitial oxygen atoms in the La-O layer, and the oxygen stoichiometry changes from 4.14 to 4.07 in the same conditions [3,12].

Temperature-programmed reduction experiments on nickelates in hydrogen have shown that interstitial oxygen ions are more weakly bonded in the structure [13]. This is consistent with the fact that the oxygen release in air starts at a much lower temperature in $La_2NiO_{4+\delta}$ compared to $La_{2-x}Sr_xNiO_{4-\delta}$ (x = 1 or 1.4).

The observation of thermal hystereses in resistivity measurements as well as in thermal expansion and neutron diffraction experiments on strontium-substituted samples seems to indicate that the establishment of oxygen equilibrium is slow and not completely reached at 2.6 °C/min heating/cooling rate. In the neutron experiment, where the sample was equilibrated in air at each temperature for at least 6 hours, the oxygen content on heating and cooling was found identical at 700°C, but not at 400°C (see oxygen site occupancy values in Table 2). As noted above about resistivity measurements, the oxygen exchange kinetics are improved at higher oxygen pressure (oxygen flow), according to expectations.

The decrease in resistance of $La_{2-x}Sr_xNiO_{4-\delta}$ samples with x = 1 or 1.4 on a heating/ cooling cycle can be understood in terms of the evolution of oxygen vacancy concentration, which is connected to the nickel oxidation state OS(Ni). At 870°C, the vacancy concentration on the O1 site is 0.13. According to Table 3, the valence distribution changes from a $Ni^{4+}/Ni^{3+}$ distribution with 40 % $Ni^{4+}$ at room temperature ($\delta = 0.04$) to a $Ni^{3+}/Ni^{2+}$ one (10% $Ni^{2+}$) at 870°C. This has important consequences on the filling and orbital degeneracy of the band orbitals with $d(x^2-y^2)$ and $d(z^2)$ parentage and may induce a local reorganization of Ni-O bonds [2, 12]. These effects are the probable cause of the anomalies observed in the temperature variation of (i) the of *a* and *c* cell parameters (Fig.4), (ii) the electrical resistivity (Fig.2). Note that this situation is quite different from that in the $La_2NiO_{4\pm\delta}$ system, where the nickel oxidation state is always lower than 3+.

## 5. Conclusions

In the $La_{2-x}Sr_xNiO_{4-\delta}$ solid solution, single phase compounds were successfully obtained for x as high as 1.4 by solid state reaction in air at temperatures up to 1300 °C. Phases with x > 1 are oxygen-deficient as prepared, and lose oxygen when heated above 500°C in air and in oxygen. We show that the high-temperature behaviour of these strontium-substituted phases differs significantly from that of the parent compound $La_2NiO_{4\pm\delta}$. This can be seen in the temperature variation of electrical resistivity, thermal expansion and cell parameters. The stoichiometry variation with temperature is confirmed, and the vacancy location in the structure, has been established by variable-temperature neutron diffraction. The high-temperature properties of these compounds are tentatively ascribed to the evolution of nickel valence at high temperature.

***Acknowledgements***

The authors gratefully acknowledge the financial support of NATO (Project CLG 982009), and the assistance of Olivier Isnard in neutron diffraction experiments at Institut Laue-Langevin, Grenoble, France.

Table1. Chemical analysis of sintered compounds $La_{2-x}Sr_xNiO_{4-\delta}$ used in this study

| x | theoretical composition | final sintering conditions | experimental oxygen content | | average Ni oxidation state |
|---|---|---|---|---|---|
| | | | *iodometric titration* | *thermogravimetry in hydrogen* | |
| 0 | $La_2NiO_4$ | 1250°C, air | 4.14 | 4.139 | +2.28 |
| 1.0 | $LaSrNiO_4$ | 1250°C, $O_2$ | 3.99 | 3.990 | +2.98 |
| 1.2 | $La_{0.8}Sr_{1.2}NiO_4$ | ibid. | 3.95 | 3.946 | +3.10 |
| 1.4 | $La_{0.6}Sr_{1.4}NiO_4$ | ibid. | 3.95 | 3.951 | +3.30 |

Table 2. Results of Rietveld refinements from neutron diffraction data on $La_{0.6}Sr_{1.4}NiO_{4-\delta}$ at different temperatures in air. $K_2NiF_4$-type structure, space group I4/mmm, λ = 1.9111 Å.

| cell parameters/Å | atom | variable coordinates | $B_{iso}$ (Å$^2$) | site occupancy | refinement parameters |
|---|---|---|---|---|---|
| ***400°C (heating)*** | | | | | |
| *a* 3.8454(2) | La,Sr | *z* = 0.3598(3) | 0.83(4) | 1 | 45 reflections |
| *c* 12.4326(6) | Ni | – | 0.49(4) | 1 | N-P+C = 2630 |
| c/a 3.233 | O1 | – | 1.19(4) | 0.981(13) | $R_p$ 3.34, $R_{wp}$ 4.74 |
| | O2 | *z* = 0.1612(5) | 1.52(4) | 1 * | $R_{exp}$ 1.21, $R_{Bragg}$ 2.94 |
| ***700°C (heating)*** | | | | | |
| *a* 3.8546(2) | La,Sr | *z* = 0.3591(4) | 1.56(4) | 1 | 46 reflections |
| *c* 12.5476(6) | Ni | – | 0.89(4) | 1 | N-P+C = 2718 |
| c/a 3.255 | O1 | – | 1.60(6) | 0.918(15) | $R_p$ 3.12, $R_{wp}$ 4.64 |
| | O2 | *z* = 0.1606(5) | 2.45(5) | 1 * | $R_{exp}$ 1.26, $R_{Bragg}$ 2.55 |
| ***870°C*** | | | | | |
| *a* 3.8545(2) | La,Sr | *z* = 0.3572(4) | 2.18(4) | 1 | 46 reflections |
| *c* 12.6626(7) | Ni | – | 1.40(3) | 1 | N-P+C = 2783 |
| c/a 3.285 | O1 | – | 1.93(5) | 0.874(14) | $R_p$ 2.66, $R_{wp}$ 3.97 |
| | O2 | *z* = 0.1613(5) | 3.10(4) | 1 * | $R_{exp}$ 1.28, $R_{Bragg}$ 2.06 |
| ***700°C (cooling)*** | | | | | |
| *a* 3.8515(2) | La,Sr | *z* = 0.3583(4) | 1.57(4) | 1 | 46 reflections |
| *c* 12.5683(8) | Ni | – | 0.83(4) | 1 | N-P+C = 2604 |
| c/a 3.263 | O1 | – | 1.44(7) | 0.918(17) | $R_p$ 3.36, $R_{wp}$ 4.99 |
| | O2 | *z* = 0.1608(5) | 2.43(5) | 1 * | $R_{exp}$ 1.33, $R_{Bragg}$ 2.69 |
| ***400°C (cooling)*** | | | | | |
| *a* 3.8434(2) | La,Sr | *z* = 0.3595(3) | 0.92(4) | 1 | 45 reflections |
| *c* 12.4440(6) | Ni | – | 0.48(4) | 1 | N-P+C = 2605 |
| c/a 3.238 | O1 | – | 1.03(5) | 0.953(17) | $R_p$ 3.49, $R_{wp}$ 4.95 |
| | O2 | *z* = 0.1602(4) | 1.62(4) | 1 * | $R_{exp}$ 3.95, $R_{Bragg}$ 2.59 |

* Refinement of this variable yielded value not significantly different from 1.

Table 3. Evolution of nickel oxidation state (OS) and interatomic distances in $La_{0.6}Sr_{1.4}NiO_{4-\delta}$ as a function of temperature in air from neutron refinement data

| T/°C | OS(Ni) | Ni-O1/Å (x4) | Ni-O2/Å (x2) |
|---|---|---|---|
| 400 | 3.32(1) | 1.923(1) | 2.004(4) |
| 700 | 3.07(1) | 1.927(1) | 2.015(5) |
| 870 | 2.90(1) | 1.927(1) | 2.042(5) |
| 700 | 3.07(1) | 1.926(1) | 2.021(6) |
| 400 | 3.21(1) | 1.922(1) | 1.993(4) |

**Figure captions**

Fig. 1. Thermogravimetric behaviour of $La_{2-x}Sr_xNiO_{4-\delta}$ with x = 1 (triangles) and x = 1.4 (squares and diamonds) in air and in oxygen.. Heating (full symbols) and cooling (open symbols) were carried out at 2.6 °C/min.

Fig. 2. Evolution of electrical resistance with temperature for ceramic samples of $LaSrNiO_{4\pm\delta}$ (1, 2) and $La_{0.6}Sr_{1.4}NiO_{4--\delta}$ (3, 4) during heating (a) and cooling (b) with rate 2.6 °C/min in air (1, 3) and in oxygen flow (2, 4). Upper panel (filled points) : data for $La_2NiO_{4\pm\delta}$.

Fig. 3. Thermal expansion of the ceramic samples $La_2NiO_{4\pm\delta}$ $La_{0.6}Sr_{1.4}NiO_{4-\delta}$ during heating and cooling with a rate 2.6 °C/min.

Fig. 4. Evolution of tetragonal cell parameters *a* and *c* with temperature (XRD data) for $La_{2-x}Sr_xNiO_{4\pm\delta}$ with x = 0 (a), x = 1 (b) and x = 1.4 (c).

Fig. 5. Structural model for $La_{2-x}Sr_xNiO_4$ compounds ($K_2NiF_4$-type structure).

Fig. 6. Observed (points) and calculated (continuous line) neutron diffraction patterns of $La_{0.6}Sr_{1.4}NiO_{4-\delta}$ at 780°C. The lower curve is the difference $I_{obs}$-$I_{calc}$. Bumps in the baseline around 25 and 52° are due to the silica tube.

Figure(s)
Click here to download high resolution image

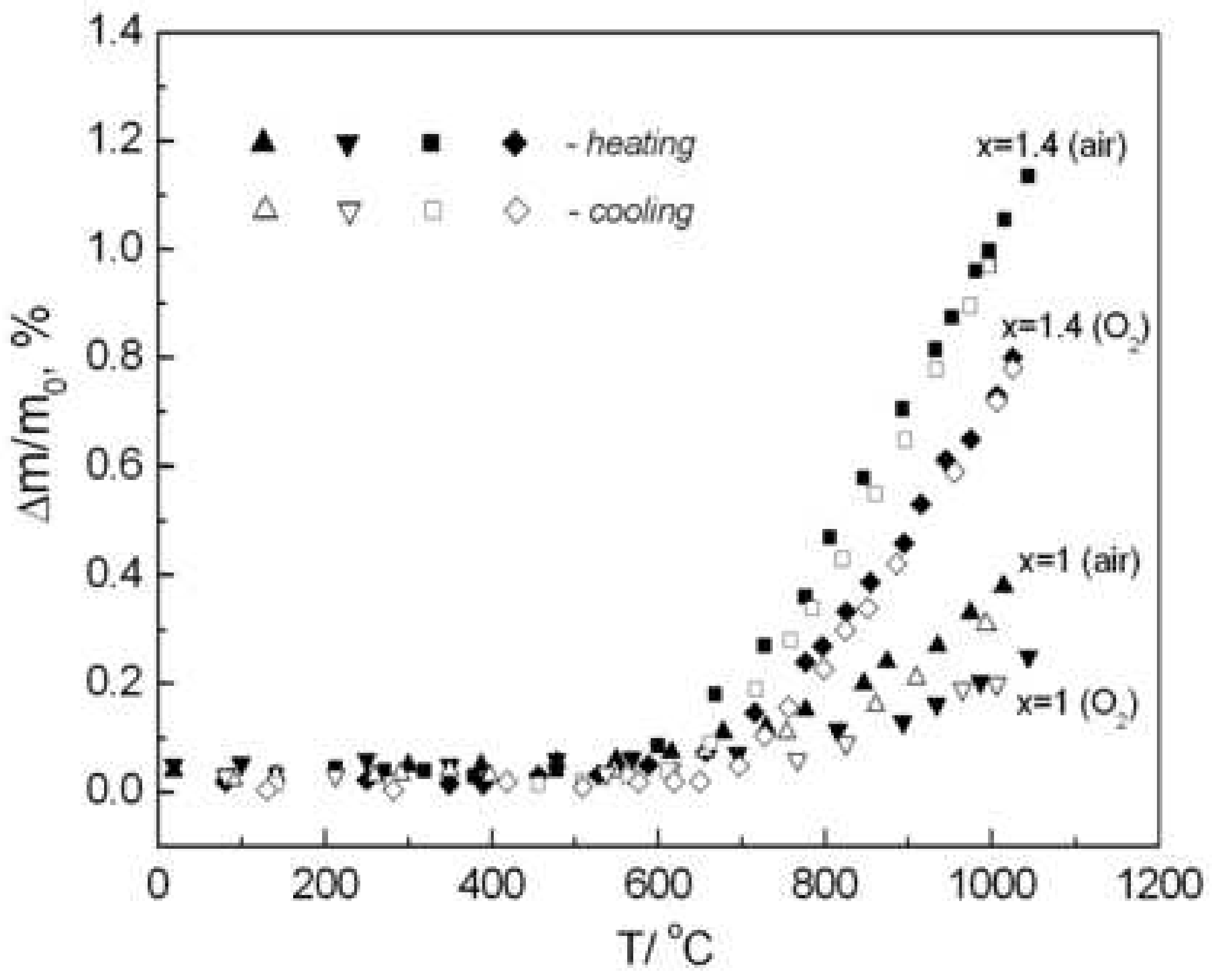

Figure(s)
Click here to download high resolution image

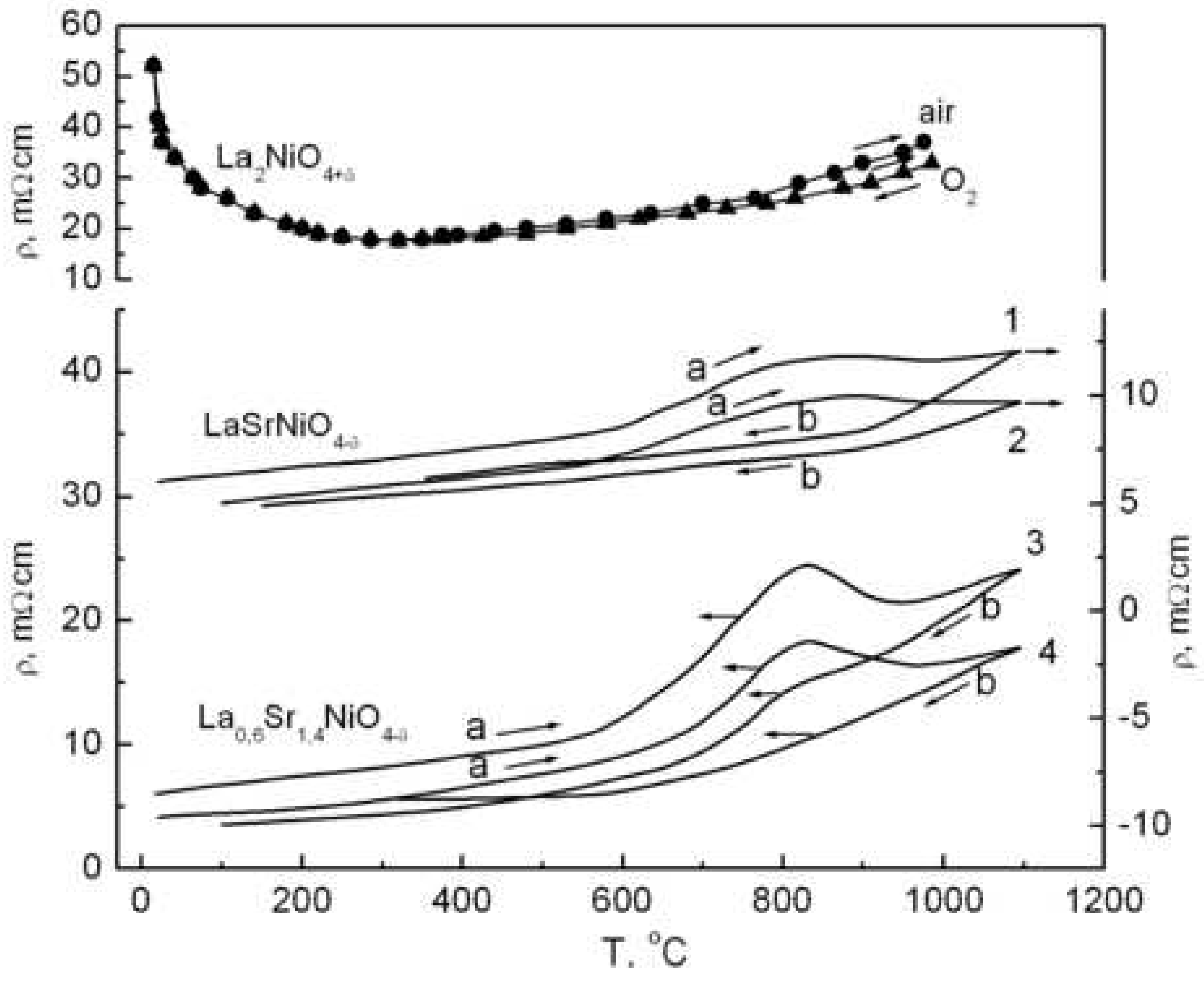

Figure(s)
Click here to download high resolution image

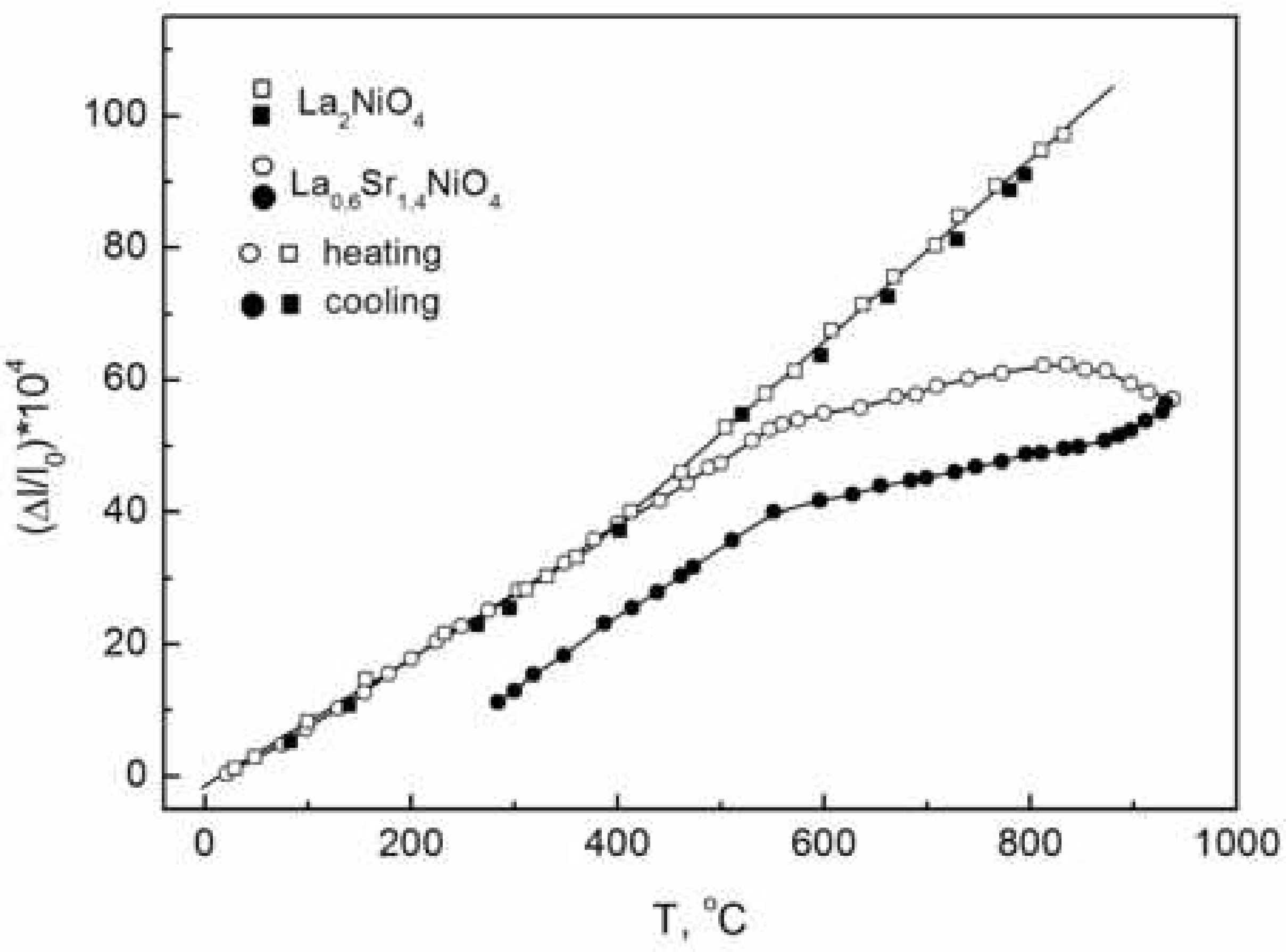

Figure(s)
Click here to download high resolution image

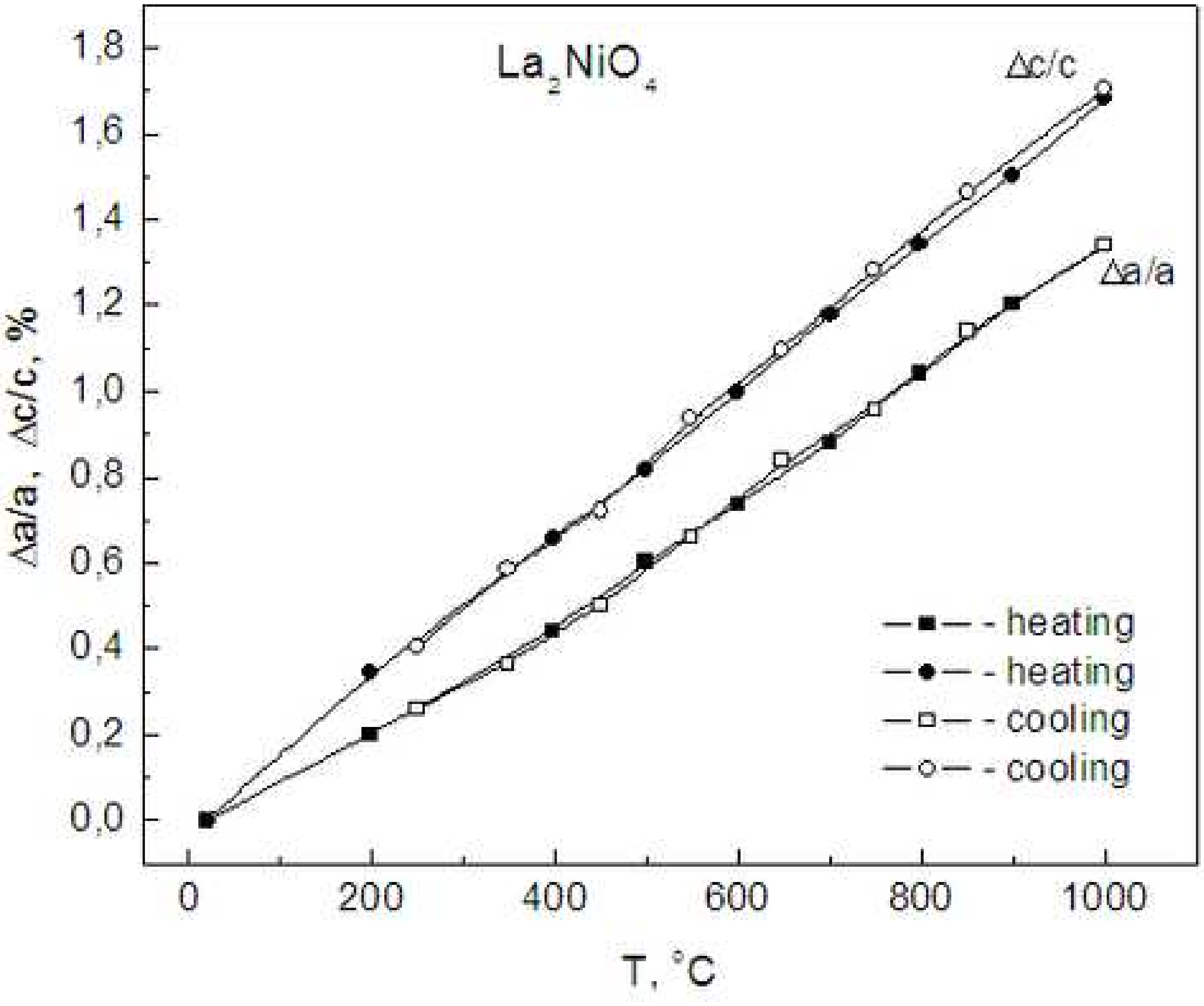

Figure(s)
Click here to download high resolution image

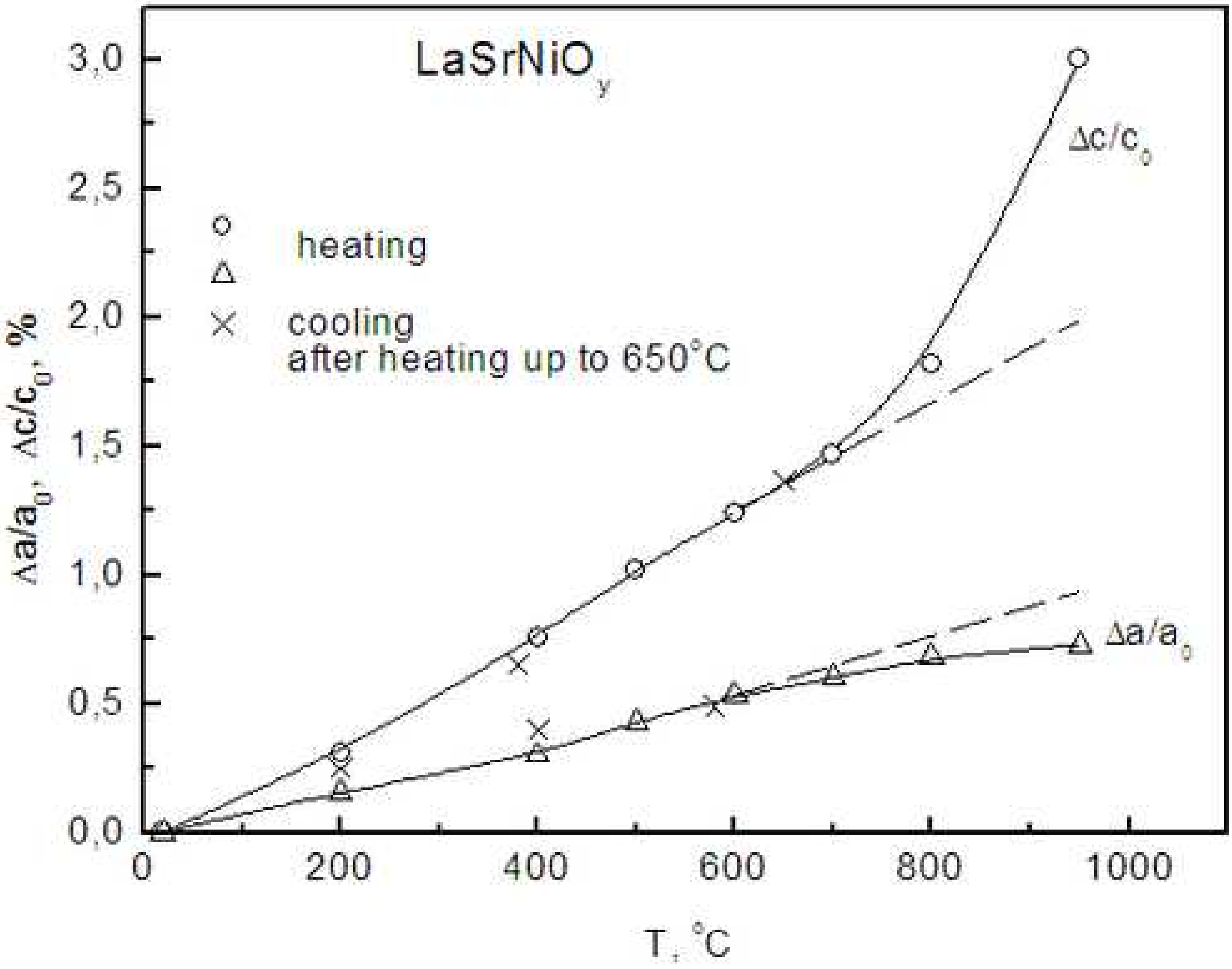

Figure(s)
Click here to download high resolution image

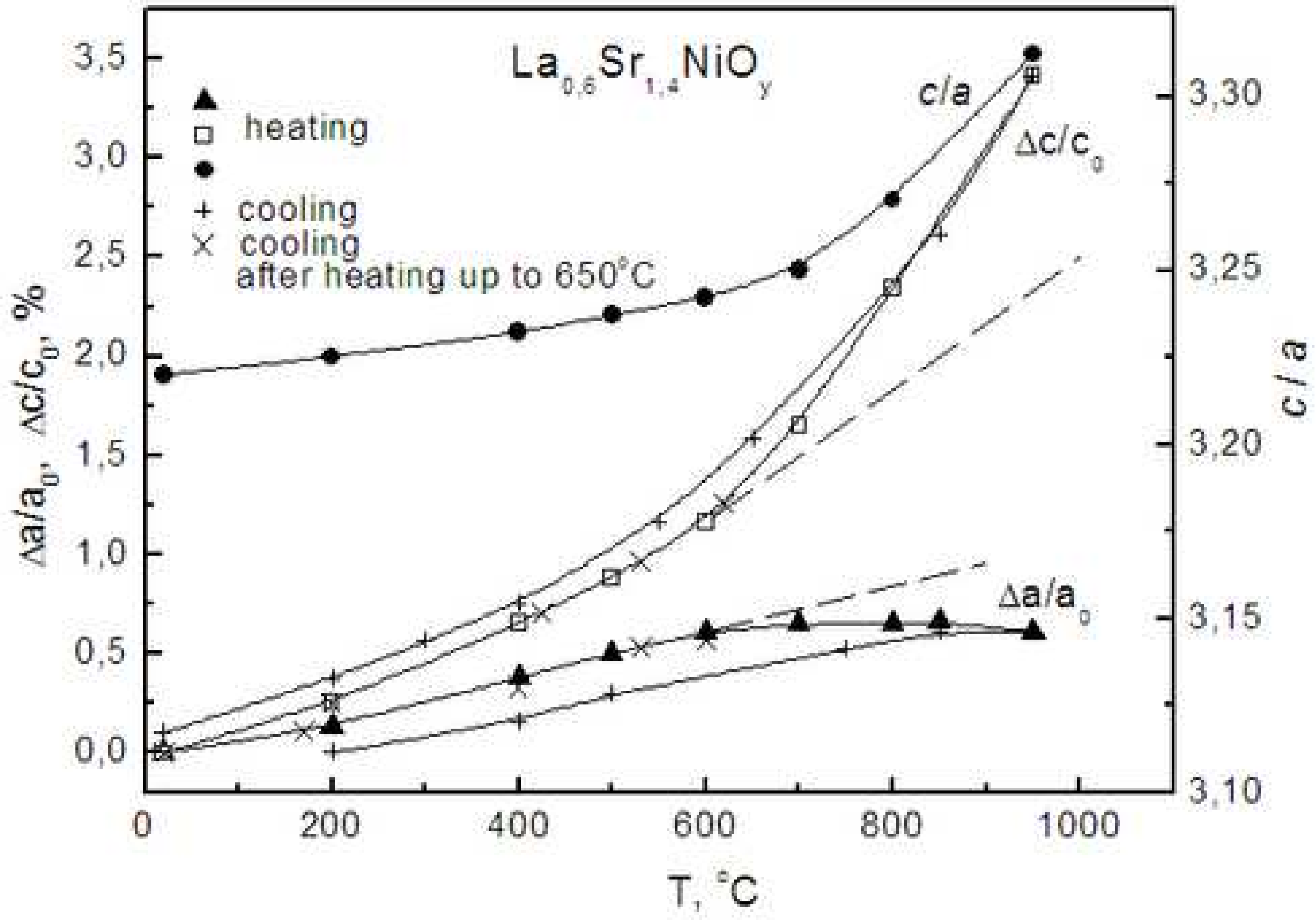

**Figure(s)**

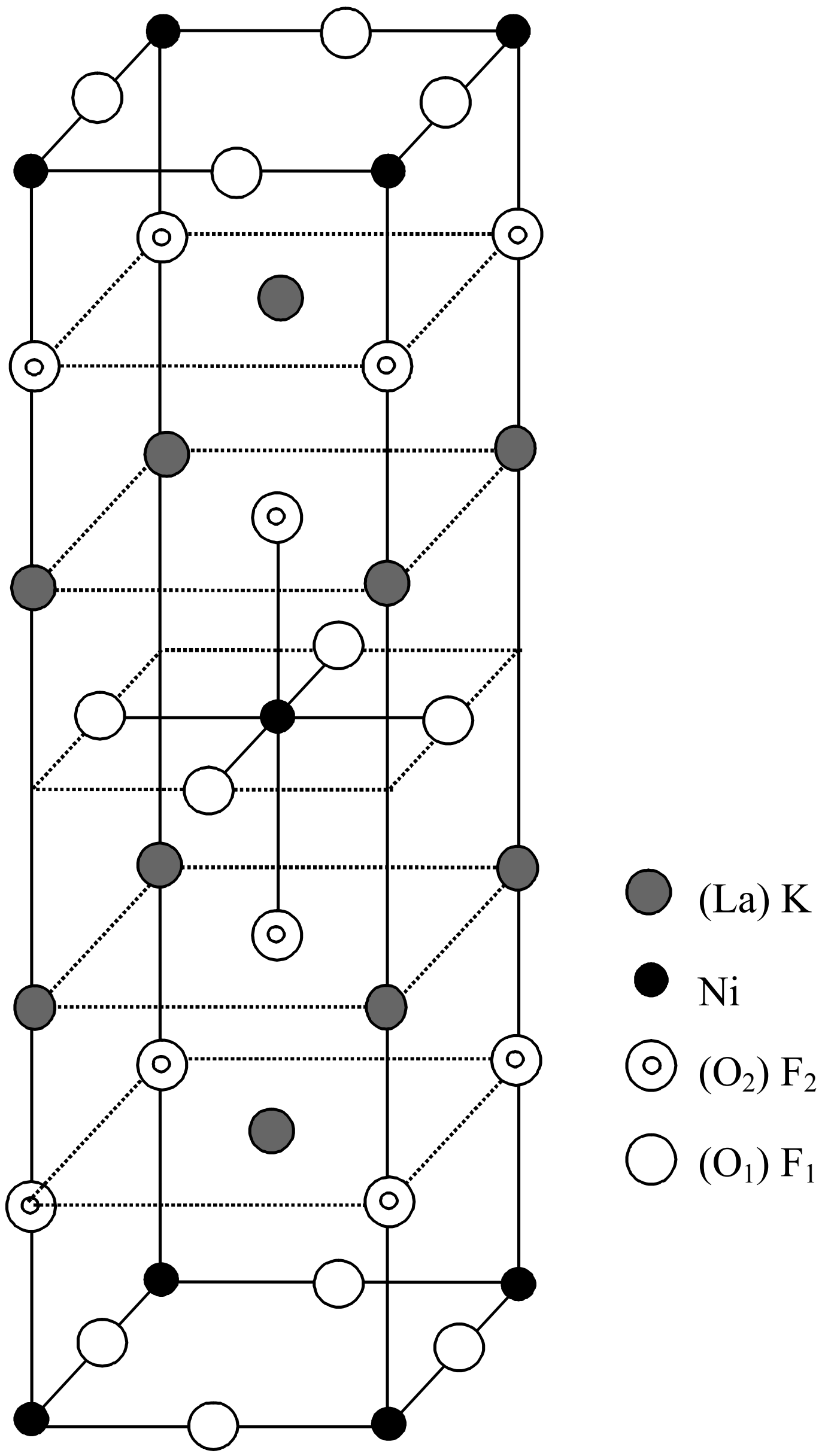

Fig.5. Structural model for $La_{2-x}Sr_xNiO_4$ compounds ($K_2NiF_4$-type structure).

Figure(s)
Click here to download high resolution image

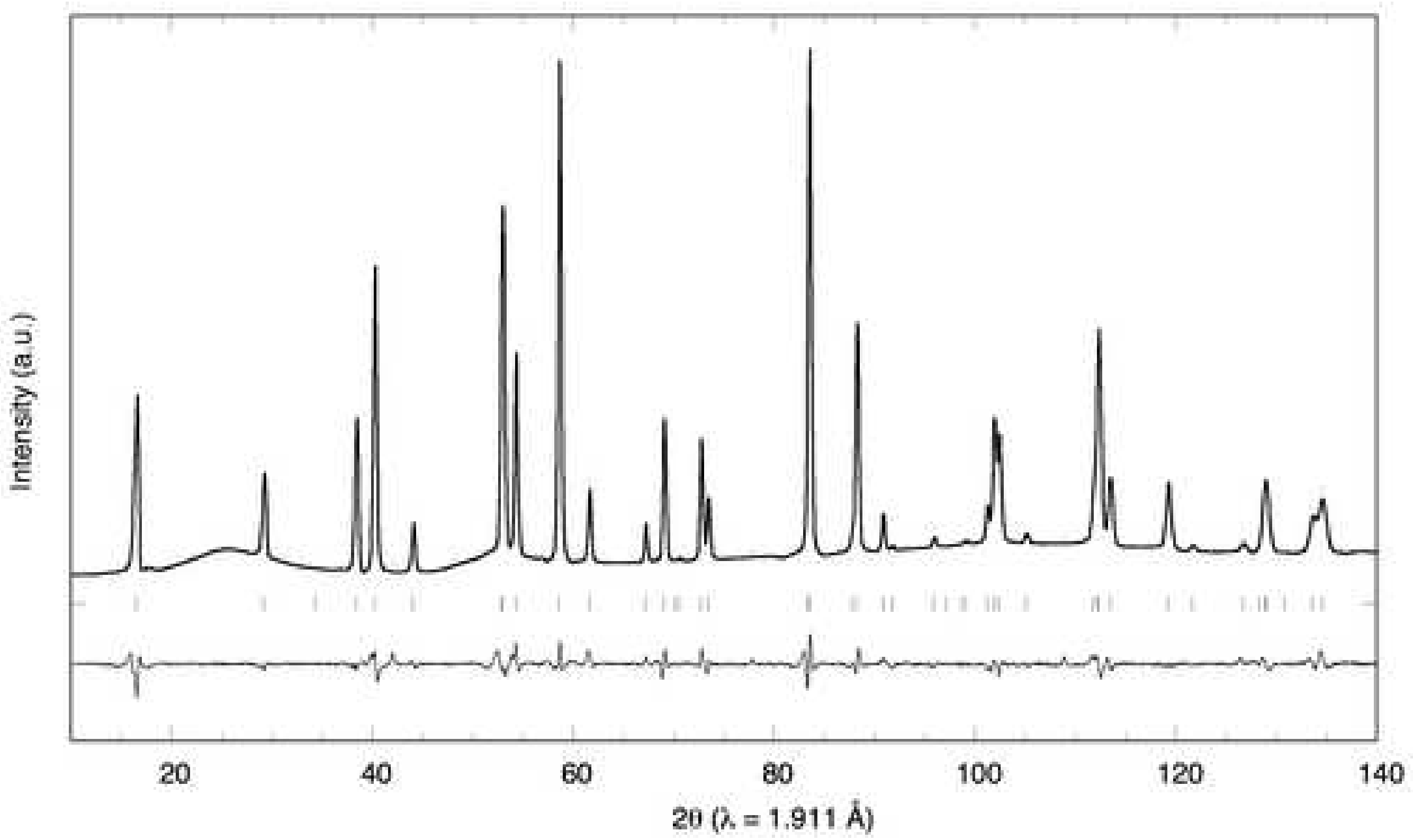